\documentclass[pmlr]{jmlr}

\usepackage{longtable}

\usepackage{booktabs}
\usepackage[load-configurations=version-1]{siunitx} 

\theorembodyfont{\upshape}
\theoremheaderfont{\scshape}
\theorempostheader{:}
\theoremsep{\newline}

\jmlrvolume{277}
\jmlryear{2025}
\jmlrworkshop{Machine Learning Meets Differential Equations: From Theory to Applications}
\usepackage[disable]{todonotes}  
\usepackage[nolist, nohyperlinks]{acronym}
\usepackage{mathtools}
\usepackage{caption}

\title[Flowing Straighter with Conditional Flow Matching for Accurate Speech Enhancement]{Flowing Straighter with Conditional Flow Matching for Accurate Speech Enhancement}

 \author{\Name{Mattias Cross} \Email{mcross2@sheffield.ac.uk}\\\and
  \Name{Anton Ragni} \Email{a.ragni@sheffield.ac.uk}\\
  \addr School of Computer Science, The University of Sheffield, Sheffield, UK}

\editors{Cecília Coelho, Bernd Zimmering, M. Fernanda P. Costa, Luís L. Ferrás, Oliver Niggemann}

\begin{document}
\begin{acronym}
\acro{cfm}[CFM]{conditional flow-matching}
\acro{sgm}[SGM]{score-based generative model}
\acro{snr}[SNR]{signal-to-noise ratio}
\acro{gan}[GAN]{generative adversarial network}
\acro{vae}[VAE]{variational autoencoder}
\acro{ddpm}[DDPM]{denoising diffusion probabilistic model}
\acro{STFT}[STFT]{short-time Fourier transform}
\acro{iSTFT}[iSTFT]{inverse short-time Fourier transform}
\acro{SDE}[SDE]{stochastic differential equation}
\acro{ODE}[ODE]{ordinary differential equation}
\acro{ou}[OU]{Ornstein-Uhlenbeck}
\acro{VE}[VE]{variance exploding}
\acro{OUVE}[OUVE]{Ornstein-Uhlenbeck process with variance exploding}
\acro{dnn}[DNN]{deep neural network}
\acro{PESQ}[PESQ]{Perceptual Evaluation of Speech Quality}
\acro{se}[SE]{speech enhancement}
\acro{tf}[T-F]{time-frequency}
\acro{elbo}[ELBO]{evidence lower bound}
\acro{WPE}{weighted prediction error}
\acro{MAC}{multiply–accumulate operation}
\acro{PSD}{power spectral density}
\acro{RIR}{room impulse response}
\acro{SNR}{signal-to-noise ratio}
\acro{LSTM}{long short-term memory}
\acro{POLQA}{Perceptual Objectve Listening Quality Analysis}
\acro{SDR}{signal-to-distortion ratio}
\acro{SI-SDR}{scale invariant signal-to-distortion ratio}
\acro{ESTOI}{Extended Short-Term Objective Intelligibility}
\acro{ELR}{early-to-late reverberation ratio}
\acro{TCN}{temporal convolutional network}
\acro{DRR}{direct-to-reverberant ratio}
\acro{nfe}[NFE]{number of function evaluations}
\acro{rtf}[RTF]{real-time factor}
\acro{MOS}[MOS]{mean opinion score}
\acro{EMA}[EMA]{exponential moving average}
\acro{SB}[SB]{Schrö\-din\-ger bridge}
\acro{SGMSE}[SGMSE]{score-based generative models for speech enhancement}
\acro{EDM}[EDM]{elucidating the design space of diffusion-based generative models}
\acro{GPU}[GPU]{graphics processing unit}
\acro{VB-DMD}[VB-DMD]{Voicebank-Demand}
\acro{SB-VE}[SB-VE]{Schrödinger bridge with variance exploding diffusion coefficient}
\acro{SB-SV}[SB-SV]{Schrödinger bridge with static variance}
\acro{ICFM}[ICFM]{independent conditional flow-matching}
\acro{SE}[SE]{speech enhancement}
\acro{FM}[FM]{flow matching}
\acro{CNF}[CNF]{continuous normalising flows}
\acro{OT}[OT]{optimal transport}
\acro{VE-SDE}[VE-SDE]{Stochastic differential equation with variance exploding diffusion coefficient}
\acro{DM}[DM]{diffusion model}
\acro{ML}[ML]{machine learning}
\acro{DDP}{direct data prediction}
\end{acronym} 
\maketitle

\begin{abstract}
Current flow-based generative speech enhancement methods learn curved probability paths which model a mapping between clean and noisy speech. 
Despite impressive performance, the implications of curved probability paths are unknown. Methods such as Schr\"odinger bridges focus on curved paths, where time-dependent gradients and variance do not promote straight paths. Findings in machine learning research suggest that straight paths, such as conditional flow matching, are easier to train and offer better generalisation. In this paper we quantify the effect of path straightness on speech enhancement quality.
We report experiments with the Schr\"odinger bridge, where we show that certain configurations lead to straighter paths. Conversely, we propose independent conditional flow-matching for speech enhancement, which models straight paths between noisy and clean speech. 
We demonstrate empirically that a time-independent variance has a greater effect on sample quality than the gradient.
Although conditional flow matching improves several speech quality metrics, it requires multiple inference steps. We rectify this with a one-step solution by inferring the trained flow-based model as if it was directly predictive.
Our work suggests that straighter time-independent probability paths improve generative speech enhancement over curved time-dependent paths.

\end{abstract}

\section{Introduction}
\label{sec:intro}
\begin{figure}[h]
    \centering
    \includegraphics[width=0.75\linewidth]{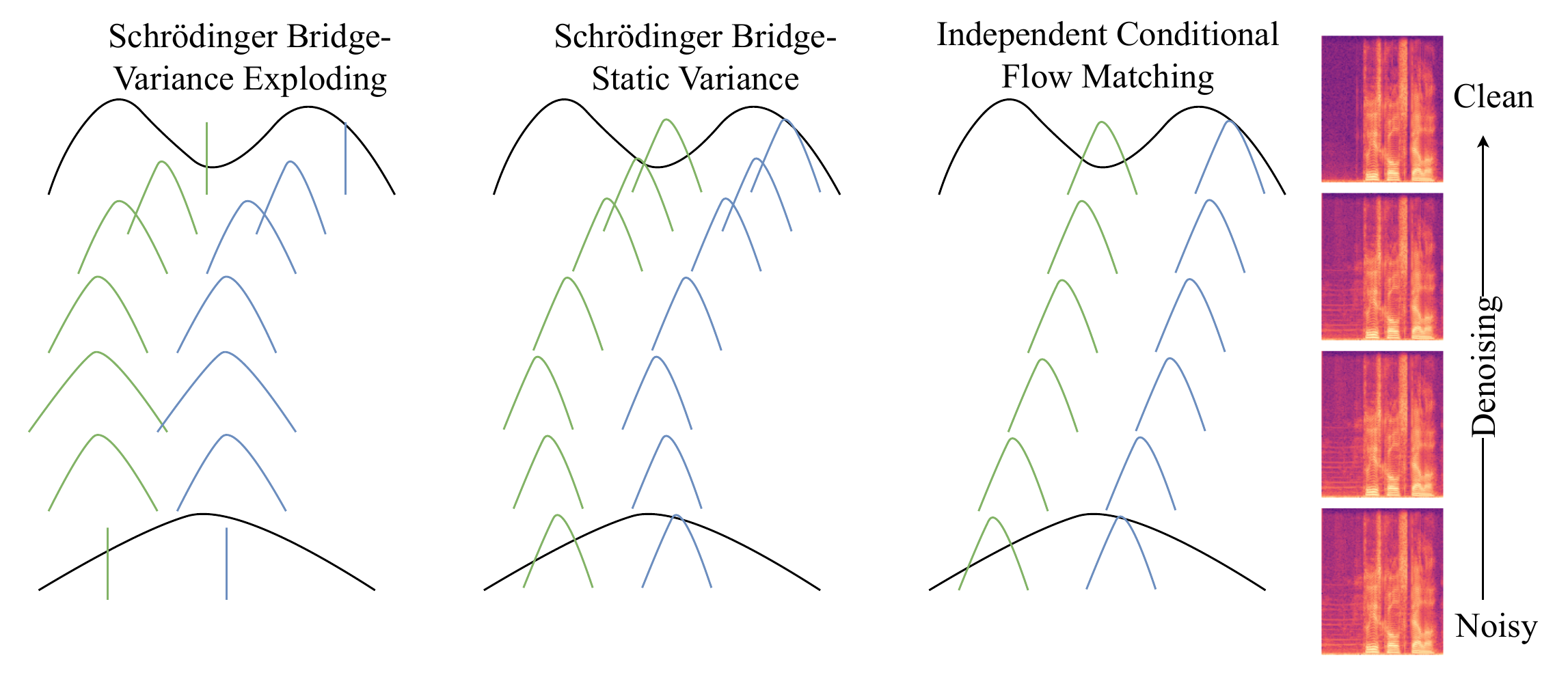}
    \caption{SB and ICFM learn a Gaussian probability path between distributions. SB is sublinear with time-varying variance that starts and ends at 0. ICFM is linear with constant variance.}
    \label{fig:sb_cfm_spec}
\end{figure}
Understanding what people say in noisy environments, such as a crowded caf\'e, is tricky for computers. Suppressing background noise in speech recordings, known as \acf{SE}, is a task that has seen many proposed solutions involving flow-based generative methods. These methods solve \ac{SE} by estimating the distribution of clean speech, which can be conditionally sampled from given noisy speech input \citep{richterInvestigatingTrainingObjectives2025}. The clean distribution is estimated with \ac{CNF}, models that learn a mapping between two distributions with a neural \ac{ODE} \citep{chenNeuralOrdinaryDifferential2019}. Neural \acp{ODE} are \acp{ODE} parameterised with a neural network to estimate a velocity field that pushes samples from a source to a target distribution, enabling a continuous mapping that can be computed with ODE solvers. The \ac{SE} problem is particularly well-suited to \acp{CNF} because samples of source-target pairs are similar: the source sample is the target with added noise and potentially reverberation. There are many methods for training \acp{CNF}: \acp{DM}~\citep{sohl-dicksteinDeepUnsupervisedLearning2015a, song2021sde}, \acfp{SB}~\citep{chen2021likelihood, debortoliDiffusionSchrodingerBridge2021, wangDeepGenerativeLearning2021}, and \ac{FM}~\citep{lipman2023flow, albergoStochasticInterpolantsUnifying2023, liuFlowStraightFast2022a, tongImprovingGeneralizingFlowbased2024}. \ac{DM} and \ac{SB} have received attention as powerful SE methods \citep{jukic2024schr, richterInvestigatingTrainingObjectives2025, richter2023speech}, with \ac{FM} being less explored at the time of writing.
The methods for training \acp{CNF} described above define a Gaussian probability path which interpolates between a pair of distributions; each method is identifiable by its probability path and source distribution (with the consistent target distribution being clean speech; Figure \ref{fig:sb_cfm_spec}). For example, \ac{SB} defines a path that solves the \ac{SB} problem between the exact noisy and clean speech distributions (elaborated in \sectionref{sec:schr}). This path is typically time-dependent, where ``time'' describes progress along the path between the pair of distributions, and where the ODE is a function of time. Since the SB path interpolates the exact data, the ODE is accurate and does not require numerous ODE steps, yielding practical inference speed. Despite the strong modelling power of SB, time-dependent gradients and variance can cause \textit{curved} paths.
Many works in the \ac{ML} literature suggest that \textit{straight} paths are preferred over curves because they are easier to train and experience less ODE sampling errors \citep{lipman2023flow, albergoStochasticInterpolantsUnifying2023, liuFlowStraightFast2022a, tongImprovingGeneralizingFlowbased2024}, giving rise of \ac{FM}. The goal of \ac{FM} is to induce straight paths by relaxing constraints on probability path design to any velocity field (flow) that interpolates source- and target-distributions.
This relaxation allows various straighter paths to be chosen, such as an optimal transport displacement interpolant \citep{mccannConvexityPrincipleInteracting1997}. This formulation produces straight paths with time-independent velocity, resulting in faster training and ODE inference, and higher sample quality \citep{lipman2023flow, lipmanFlowMatchingGuide2024}. The intuition behind this is that straighter paths are easier to sample with ODE solvers, and fewer curves demand less modelling power from the neural ODE. However,  the originally proposed FM is not well-suited for data-to-data tasks such as SE because it  considers paths from the standard normal distribution, not empirical data such as the noisy speech distribution. To relax this, independent conditional flow-matching (ICFM) generalises FM to the independent coupling of two general distributions, e.g. a path between paired data \citep{tongImprovingGeneralizingFlowbased2024}.
Although SB produces state-of-the-art results for SE, it is unknown if its time-dependent and potentially curved path could be improved by using straighter paths. Further, time-independent models such as ICFM have not been proposed for direct SE, although there has been work on ICFM from audio-visual embeddings for SE \citep{jungFlowAVSEEfficientAudioVisual2024}. 
Concurrent to our work, FM with time-varying variance has been adapted to the SE task in FlowSE \citep{leeFlowSEFlowMatchingbased2025}, and an alternative time-varying FM set-up with modifications for improved one-step performance has also been proposed \citep{korostikModifyingFlowMatching2025}. Although these two works are relevant, neither explores time-independent variance.
In light of this, we explore the impact of time-dependence on the probability path for SE by comparing SB to ICFM (\figureref{fig:sb_cfm_spec}). We show that although certain configurations of SB ensue time-independent gradients, a time-independent variance is not supported.  
To identify the significance of time-independent gradient and variance on sample quality, \textbf{we propose \ac{SB-SV}, a model whose gradient is equal to SB but with time-independent variance}.
As an example of a model which, by design, has time-independent paths, \textbf{we propose and evaluate a novel formulation of independent conditional flow-matching (ICFM) for \ac{SE}}. 
We find that speech quality metrics increase when introducing SB-SV, which are then further improved with ICFM.
These observations suggest that time independence is important for high sample quality.
We also evaluate the link between the number of ODE steps and speech quality. SB is robust to one-step ODE inference, but our proposed models require thirty steps to achieve the best results. We rectify this by \textbf{proposing a simple approach for one-step inference with \ac{DDP} of clean speech from noisy speech input}. We find samples from DDP to be on par, if not surpass, those produced by ODEs.

The rest of this paper outlines the SB method along with our proposed SB-SV and ICFM for SE, including our DDP inference in  \sectionref{sec:method}. \sectionref{sec:experiments} details experiments. Finally, the results are presented with a discussion in \sectionref{sec:results_and_disc}.

\section{Flow-based models for speech enhancement}
\label{sec:method}
\subsection{General definition}
\label{sec:fbm}
In generative SE, flow-based models are defined as models that learn a marginal path $p_t$ between a prior $p_1$ and the clean speech distribution $p_0$. A Gaussian probability path $p_t$ that satisfies these boundaries can be defined by 
\begin{equation}
\label{eq:perturbation-kernel}
    p_{t}(\mathbf x_t|\mathbf x_0, \mathbf y) \coloneq \mathcal{N}_\mathbb{C}(\mathbf x_t; \boldsymbol\mu_t(\mathbf x_0, \mathbf y), \sigma_{\mathbf x_t}^2 \mathbf{I})\,,
\end{equation}
where $\mathbf{x}_t\!\in\!\mathbb{C}^d$ is the process state at time $t \!\in\! \left[0, 1 \right]$ and $\mathbf y\!\in\!\mathbb{C}^d$ is a noisy speech sample, and $\mathbf x_0\sim p_0$ is clean speech; $\mathbb{C}^d$ is the complex \ac{STFT} domain. The prior $p_1$, mean $\boldsymbol{\mu}_t$, and variance $\sigma_{\mathbf x_t}^2$ are not arbitrary and must be defined during model design (later described in \equationref{eq:sb_mean_variance,eq:icfm_mean_variance,eq:sbsv_mean_variance}). For example, \ac{SGMSE} \citep{welker2022speech} and \ac{ICFM} define $p_1$ as a Gaussian distribution centred around $\mathbf{y}$, and SB defines $p_1$ as the exact noisy data distribution with samples $\mathbf{y}$. When computing the path $p_t$ on new data, the clean speech $\mathbf x_0$ is unknown, leaving $p_t$ intractable. Flow-based models aim to train a neural ODE to estimate $p_t$ without requiring $\mathbf x_0$. This is achieved by training a neural network $F_\theta$  to predict the gradient of $p_t$.
For a given discretisation schedule $(t_N=1, t_{N-1}, \dots, t_0=0)$ with $N$ steps, the neural ODE sampler is
\begin{equation}
    \label{eq:ode}
    \mathbf x_{t_{n-1}} = a_{n} \mathbf x_{t_{n}} + b_{n} F_\theta(\mathbf x_{t_{n}}, \mathbf y, t_n) + c_{n} \mathbf y, \quad \mathbf x_{t_{N}} = \mathbf y
\end{equation}
where $a_n$, $b_n$, and $c_n$ are determined according to the designed path~\eqref{eq:perturbation-kernel}, shown in \equationref{eq:sb_ode,eq:fm_ode,eq:fm_ode2} below. The rest of this section outlines examples of the variety of paths possible with this framework, specifically paths of varying straightness.

\begin{figure}[t]
    \centering
    \includegraphics[width=0.5\linewidth]{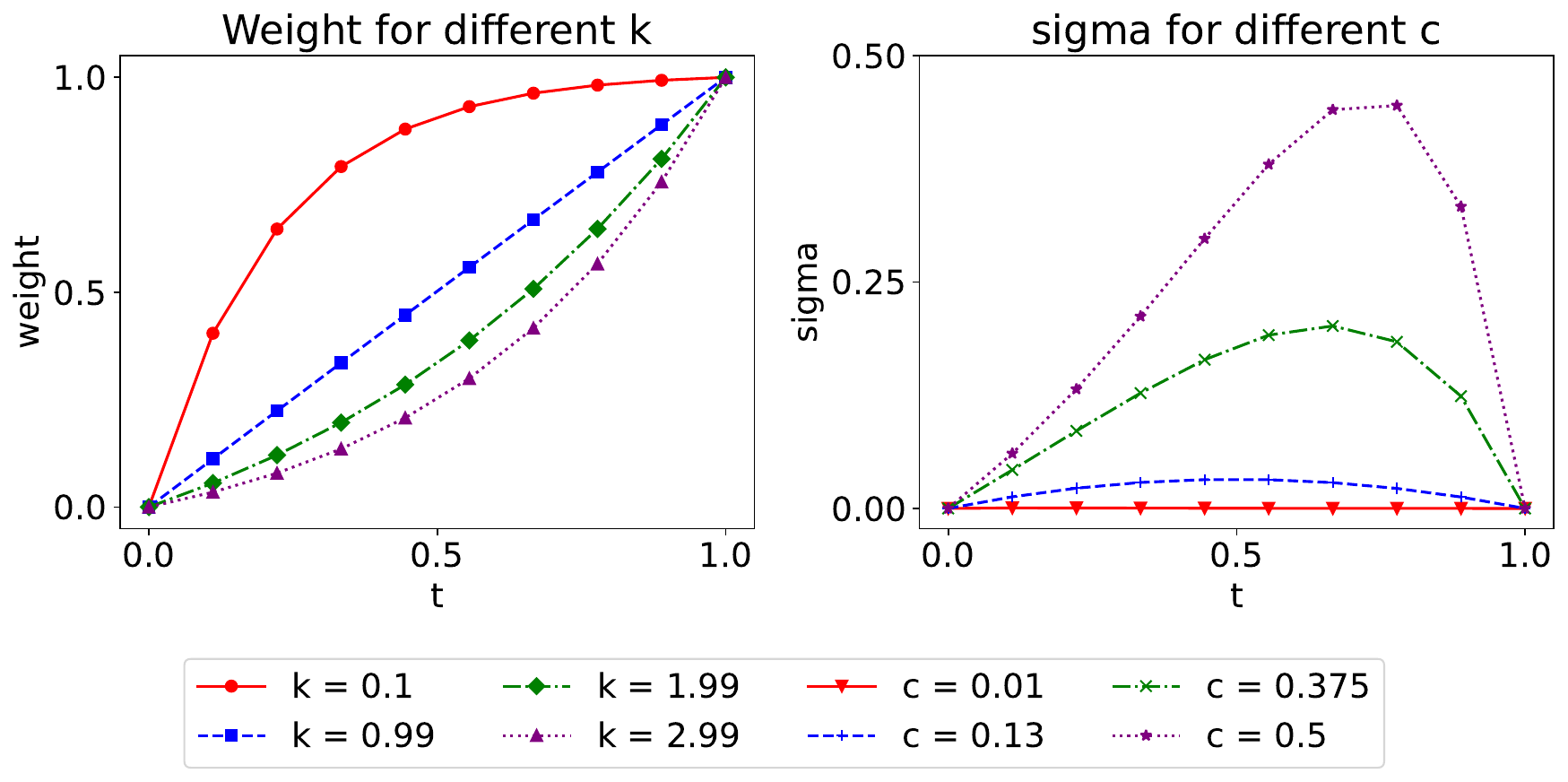}
    \caption{SB is defined by $k$ and $c$. The parameter $k$ defines the base of the interpolation weight, and $c$ the variance scale \eqref{eq:sb_mean_variance}. SB is most linear when $k=0.99$}
    \label{fig:k_c}
\end{figure}

\subsection{\acf{SB-VE}}
\label{sec:schr}
The SB problem originally considers a group of particles that we assume to move via Brownian motion, with position distribution observations $p_x$ and $p_y$ at times 0 and 1 respectively \citep{schrodinger1932theorie, leonardSurveySchrOdinger2013}. Then, imagine an unexpected (rare) event occurs such that our observation at time 1 differs substantially from what would be predicted by Brownian motion. The SB problem lies in finding the most likely path between our two observations that adheres most to Brownian motion. Formally, SB is defined as finding the probability path $p$ between boundaries $p_x$ and $p_y$ that minimises the Kullback-Leibler divergence $D_\text{KL}$ w.r.t. a pre-specified Brownian reference $p_\text{ref}$
\begin{equation}
    \min _{p \in \mathcal{P}_{\left[0, 1\right]}} D_{\text{KL}}(p, p_{ref}) \quad s.t. \quad p_0=p_x, p_1=p_y
    \label{eq:sb_kld}
\end{equation}
where $\mathcal{P}_{\left[0, 1\right]}$ is the space of all probability paths between $t=[0,1]$.
Many works in the \ac{ML} community use the SB problem to model  exact distribution-to-distribution processes that flow similarly to diffusion \citep{debortoliDiffusionSchrodingerBridge2021, chen2021likelihood, vargasMachinelearningApproachesEmpirical2021}.
The diffusion-based SGMSE uses Brownian motion, but a prior mismatch is introduced because the noisy speech distribution cannot be accurately represented by Brownian motion~(\cite{layReducingPriorMismatch2023}. Therefore, SB approaches for SE \citep{jukic2024schr, wang2024diffusion} allow a \ac{DM} to be trained that respects the boundary conditions between noisy and clean speech distributions. To solve the SB-problem, one can use a closed-form solution between Gaussian measures, such as $p_0$ and $p_1$ \citep{bunneSchrodingerBridgeGaussian2023}. We follow prior works \citep{jukic2024schr, richterInvestigatingTrainingObjectives2025}  and solve the SB problem between noisy and clean speech data with a \textit{stochastic differential equation with a variance-exploding diffusion coefficient} \citep{song2021sde} as a Brownian reference  
\begin{equation}
  \label{eq:sb_mean_variance}
  \boldsymbol{\mu}_t(\mathbf x_0, \mathbf y) = \left(1 - \frac{\sigma_t^2}{\sigma_1^2}\right) \mathbf{x}_0 + \frac{\sigma_t^2}{\sigma_1^2} \mathbf{y} ,
  \quad
  \sigma_{\mathbf x_t}^2 = \sigma_t^2 \left(1 - \frac{\sigma_t^2}{\sigma_1^2}\right), \quad \sigma_t^2 = \frac{c(k^{2t}-1)}{2 \log k}.
\end{equation}
The hyperparameters $c$ and $k$ change the shape of the probability path. \figureref{fig:k_c}
shows how these values affect the interpolation weight $\frac{\sigma_t^2}{\sigma_1^2}$ and the variance $\sigma_{\mathbf x_0}^2$. Typical values are $k=2.6$ and $c=0.4$ \citep{jukic2024schr}, which results in sub-linear interpolation between the data boundaries with exponentially increasing variance satisfying $\sigma_{\mathbf x_0}^2 = \sigma_{\mathbf x_1}^2 = 0$. It can be seen that the gradient and variance of this path are time-dependent, but the gradient becomes more linear as $k\rightarrow1$ (\figureref{fig:k_c}). As stated in \sectionref{sec:fbm}, clean data samples $\mathbf x_0$ are unknown during inference, which motivates training a neural network $F_\theta$ to estimate the clean data given the current sample along the probability path
\begin{equation}
    \mathcal{L}_\text{DP} \coloneq \lVert F_\theta(\mathbf x_t, \mathbf y, t) - \mathbf x_0 \rVert^2_2,
    \label{eq:dp_loss}
\end{equation}
where $\mathbf y$ and $\mathbf x_0$ are sampled from paired data, $t \sim \mathcal{U}[0,1]$, and $\mathbf x_t \sim p_t(\mathbf x_t | \mathbf x_0, \mathbf y)$. This data prediction allows the gradient of $p_t$ to be indirectly calculated and sampled with an SB ODE defined by \cite{jukic2024schr} as
\begin{equation}
    \label{eq:sb_ode}
    a_{n} = \frac{\sigma_{t_{n-1}} \bar{\sigma}_{t_{n-1}}}{\sigma_{t_{n}} \bar{\sigma}_{t_{n}}}, \quad
    b_{n} = \frac{1}{\sigma_1^2} \left( \bar{\sigma}_{t_{n-1}}^2 - \frac{\bar{\sigma}_{t_{n}} \sigma_{t_{n-1}} \bar{\sigma}_{t_{n-1}}}{\sigma_{t_{n}}} \right), \quad
    c_n = \frac{1}{\sigma_1^2} \left( \sigma_{t_{n-1}}^2 - \frac{\sigma_{t_{n}} \sigma_{t_{n-1}} \bar{\sigma}_{t_{n-1}}}{\bar{\sigma}_{t_{n}}} \right), 
\end{equation}
where $\bar{\sigma}_t=\sigma_1 - \sigma_t$.
The above allows us to predict clean speech from noisy speech by solving the SB ODE \eqref{eq:ode}. Not only are the gradient and variance time-dependent, but the ODE solver is also time-dependent.

\subsection{Independent
conditional flow-matching (ICFM)}
\label{sec:icfm}
ML research suggests that FM is a good form of flow-based model because straight paths are easier to learn and result in fewer ODE errors, improving sample quality \citep{liuFlowStraightFast2022a, albergoStochasticInterpolantsUnifying2023, lipman2023flow}. Here, we outline our first proposed model as a method to train ICFM for the SE task. As described in \sectionref{sec:intro}, ICFM is a generalisation of FM which considers the optimal path between independently coupled distributions. This is generally defined as McCann's interpolation \citep{mccannConvexityPrincipleInteracting1997}, which we write as a probability path for SE 
\begin{equation}
  \label{eq:icfm_mean_variance}
  \boldsymbol{\mu}_t(\mathbf x_0, \mathbf y) \coloneq (1 - t) \mathbf{x}_0 + t \mathbf{y} ,
  \quad
  \sigma_{\mathbf x_t}^2 \coloneq c,
\end{equation}
where $c$ is a hyperparameter controlling variance. 
As seen in the SB probability path \eqref{eq:sb_mean_variance}, the clean speech sample $\mathbf x_0$ is yet again unknown during inference, requiring a model trained with the data prediction loss \eqref{eq:dp_loss}. A trained neural data predictor can then be used as a neural ODE \eqref{eq:sb_ode} with the following coefficients
 \begin{equation}
    \label{eq:fm_ode}
    a_{n} = 1,  \quad
    b_{n} = \frac{1}{N}, \quad
    c_n = -\frac{1}{N}. 
\end{equation}
Compared to the SB, the gradient and variance of the ICFM probability path ($\mathbf x_0 - \mathbf y$ and $c$ respectively) do not depend on $t$; the path is straight (time-independent). 
Contrary to SB, which uses a data prediction loss, we can directly learn the gradient of the probability path with an FM loss
\begin{equation}
    \mathcal{L}_\text{FM} \coloneq \lVert F_\theta(\mathbf x_t, \mathbf y, t) - (\mathbf x_0  - \mathbf y)\rVert^2_2,
    \label{eq:fm_loss}
\end{equation}
which can be sampled with
 \begin{equation}
    \label{eq:fm_ode2}
    a_{n} = 1,  \quad
    b_{n} = \frac{1}{N}, \quad
    c_n = 0. 
\end{equation}
It can be shown that ICFM is a path between a Gaussian convolution over the exact data boundaries (proposition 3.3 from \cite{tongImprovingGeneralizingFlowbased2024}), contradicting the exact data interpolation SB provides. This means there is added variance (noise) to the boundaries, which may cause inaccurate predictions but may also help with regularisation.

\subsection{\acf{SB-SV}}
\label{sec:sbsv}
Up to this point, we have discussed two approaches for straighter paths: a special case of SB-VE has straighter gradients ($k=0.99$), and ICFM additionally has time-independent variance. Neither has solely time-independent variance, leading to our second proposed model: \acf{SB-SV}. SB-SV is an example of a path with a time-dependent gradient from \eqref{eq:sb_mean_variance} with a time-independent variance from \eqref{eq:icfm_mean_variance}. We define the SB-SV path as
\begin{equation}
  \label{eq:sbsv_mean_variance}
  \boldsymbol{\mu}_t(\mathbf x_0, \mathbf y) \coloneq \left(1 - \frac{\sigma_t^2}{\sigma_1^2}\right) \mathbf{x}_0 + \frac{\sigma_t^2}{\sigma_1^2} \mathbf{y} ,
  \quad
  \sigma_{\mathbf x_t}^2 \coloneq c,
\end{equation}
where $\sigma_t$ is defined as the same as in \eqref{eq:sb_mean_variance}. SB-SV is trained with the data prediction loss \eqref{eq:dp_loss} and sampled with the SB ODE \eqref{eq:sb_ode}. Since $\sigma_{\mathbf x_t}^2$ never reaches zero at the boundaries, it no longer satisfies the boundary conditions of the SB problem, so it must be seen as a \textit{modified SB model} whose mean solves the SB problem, but its variance does not. Trading exact data interpolation for time-independent variance may lead to a model that is easier to sample, but risks both a prior and target distribution mismatch due to the variance assigned at the boundaries of the probability path.  Although a static variance promotes straighter paths, the variance added to the target distribution may increase the number of ODE steps required to overcome the error introduced by the variance. The impacts of this prior mismatch and added variance are reported later in \sectionref{sec:results_and_disc}.

\subsection{Inference with \acf{DDP}}
\label{sec:1step}
To avoid such multi-step inference with ODE solvers, we propose a formula that exploits the data predictive properties of flow-based models to extract the clean speech data $\mathbf x_0$ directly from noisy input $\mathbf y$.
Given that models trained with the data prediction loss \eqref{eq:dp_loss} predict data, clean speech can be sampled in one step with
\begin{equation}
    \mathbf x_0 \coloneq F_\theta(\mathbf y, \mathbf y, 1),
\end{equation}
 and models trained with FM \eqref{eq:icfm_mean_variance}  predict a gradient towards clean data ($\mathbf x_0 - \mathbf y$), so we add $\mathbf y$ to the model output
\begin{equation}
    \mathbf x_0 \coloneq F_\theta(\mathbf y, \mathbf y, 1) + \mathbf y.
\end{equation}
The above formulae provide a one-step method for clean speech prediction that does not require \ac{ODE} solvers for all $t$. 

\section{Experimental Setup}
\label{sec:experiments}
To survey the advantages of straighter probability paths, we investigate time-independent gradients and variance. We evaluate our proposed methods, SB-SV (\sectionref{sec:sbsv}), ICFM (\sectionref{sec:icfm}), and baseline SB-VE (\sectionref{sec:schr}). SB-VE and SB-SV have time-dependent gradient and time-independent variance, respectively, and their gradients straighten as \mbox{$k \rightarrow 1$}.  Specifically, we train SB-VE and SB-SV with $k=2.6$ and $k=0.99$ to compare the significance of a straighter gradient. Then, we employ ICFM with both DP \eqref{eq:dp_loss} and FM \eqref{eq:fm_loss} loss. As stated in \sectionref{sec:icfm}, ICFM has both time-independent gradient and variance, promoting straighter paths. For inference, we use the Euler method as an ODE solver, ranging from 1 to 50 steps, and compare with our proposed \ac{DDP} method (\sectionref{sec:1step}).

\subsection{Metrics}
Standard practice measures speech quality with intrusive and non-intrusive metrics. For intrusive \ac{SE} metrics, we measure PESQ~\citep{rixPerceptualEvaluationSpeech2001} for predicting speech quality, ESTOI~\citep{jensen2016algorithm} as a measure of speech intelligibility and \ac{SI-SDR}~\citep{leroux2018sdr} measured in dB. We also measure non-intrusive metrics that predict quality from the predicted clean speech alone. Firstly, we compute the common metric DNSMOS~\citep{reddy2021dnsmos},\footnote{https://github.com/microsoft/DNS-Challenge/tree/master/DNSMOS} which employs a neural network trained on human ratings (\ac{MOS}).  Secondly, we use WhiSQA, a non-intrusive \ac{MOS} prediction network shown to correlate well with human judgment \citep{closeHallucinationPerceptualMetricDriven2024,close2025whisqanonintrusivespeechquality}.\footnote{https://github.com/leto19/WhiSQA} All of the above metrics score higher for better quality speech.

\subsection{Model, baseline, and data}
Following \cite{jukic2024schr}, we train all models until validation SI-SDR converges, then choose the checkpoint with the best validation PESQ. Unless stated, we run \ac{ODE} samplers for 50 steps, with batch size 8 and the same STFT settings as \cite{richterInvestigatingTrainingObjectives2025}.

The neural estimator $F_\theta$ employs the NCSN++ architecture \citep{song2021sde} using the same parameterisation described in \cite{richter2023speech}. All experiments use the time-domain auxiliary loss \citep{jukic2024schr}.
We release our code and speech samples,\footnote{https://github.com/Mattias421/cfmse} which build off the repository from \cite{richterInvestigatingTrainingObjectives2025}. 
As a baseline, we use SB-VE \citep{jukic2024schr} trained with our settings above. We train and test all experiments on the \ac{VB-DMD} dataset~\citep{valentini2016investigating}, a common benchmark for \ac{SE} containing clean speech recordings from 28 speakers with added background noise, e.g. caf\'e, traffic. We use speakers p226 and p287 for validation. Non-intrusive evaluation of the clean speech yields 3.53 DNSMOS and 4.53 WhiSQA. 
\section{Results and discussion}
\label{sec:results_and_disc}
\begin{figure*} 
    \begin{minipage}[t]{0.55\textwidth} 
        \vspace{0pt}
        \centering
        \vspace{\abovecaptionskip} 
        \resizebox{\linewidth}{!}{
        \begin{tabular}{cllll|ccccc}
        \toprule
               Path& Loss&Inference& k&$c$&  PESQ &  ESTOI &  SI-SDR&   DNSMOS&WhiSQA \\
    \midrule
     Noisy& -&-&-&-& 1.97& 0.79& 8.4&  3.05&3.11\\
             \midrule
    SB-VE& DP&ODE& 2.6&0.4& 2.92& 0.87& 19.3& 3.56 & 4.46\\
     SB-VE&  DP&ODE& 0.99& 0.375& 2.92 & 0.88 & 19.5& 3.56 &4.47 \\
             \midrule
    SB-SV& DP&ODE& 2.6&0.15& 2.98 & 0.88 & 19.4 & 3.58   & \textbf{4.51   }\\
    SB-SV& DP&ODE& 0.99&0.1& 2.86 & 0.88 & 19.5 & 3.58   & 4.47   \\
    ICFM& DP&ODE& -&0.1& 2.98 & 0.88 & 20.1 & 3.59  & 4.49\\
    ICFM& FM&ODE& -&0.1& 2.91 & 0.88 & 20.3 & \textbf{3.60} & 4.50 \\
    \midrule
             SB-VE   & DP&DDP& 2.6&0.4&  2.92&  0.87&  19.4&   3.55&4.45\\
             SB-SV& DP&DDP& 2.6&0.15& 2.98& 0.88& 19.9&  3.58&4.50\\
             SB-SV& DP&DDP& 0.99&0.1& 2.99& 0.88& 20.0&  3.57&4.50\\
             ICFM& DP&DDP& -&0.1&  \textbf{3.05}&  0.88&  20.2&   3.58&\textbf{4.51}\\
            ICFM& FM&DDP& -&0.1&  3.00&  0.88&  \textbf{20.4}&   3.59&\textbf{4.51}\\
            \bottomrule
        \end{tabular}
        } 
        \captionof{table}{Mean speech quality metrics on \ac{VB-DMD} of our SB-SV and ICFM with FM loss and \ac{DDP} inference over SB-VE \citep{jukic2024schr} baseline. $k=0.99$ induces straightness and $c$ scales variance.}
        \label{tab:results}
    \end{minipage}%
    \hfill 
    \begin{minipage}[t]{0.44\textwidth} 
        \vspace{0pt} 
        \centering
        \includegraphics[width=\linewidth]{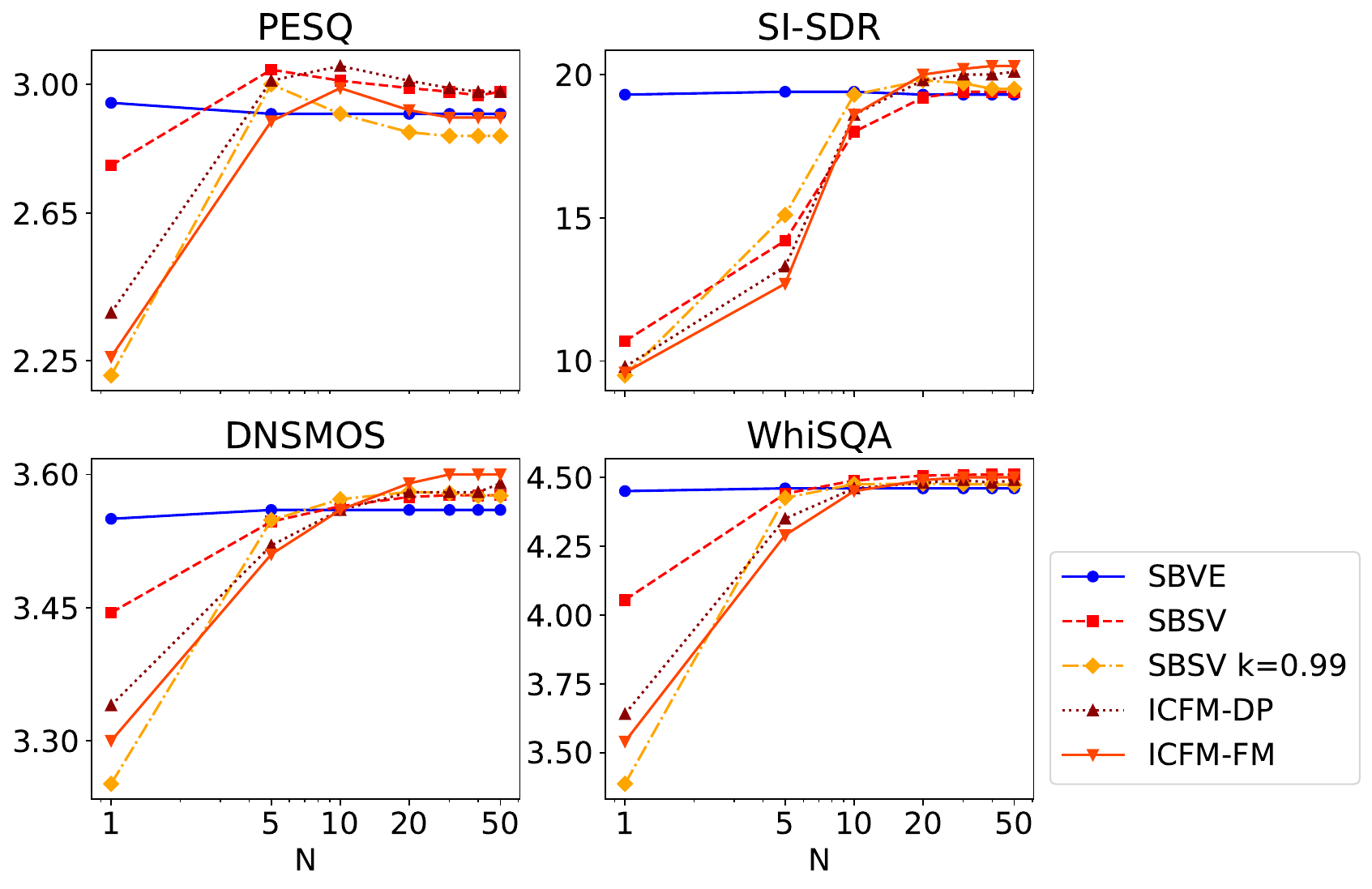} 
        \captionof{figure}{Comparing the baseline SB-VE with proposed models over various ODE steps $N$. }
        \label{fig:N}
    \end{minipage}
\end{figure*}
Our results are displayed in \tableref{tab:results}. Our proposed straighter paths \ac{SB-SV} (\sectionref{sec:sbsv}) and \ac{ICFM} (\sectionref{sec:icfm}) suggest improved speech quality across all metrics over the curved SB-VE (\sectionref{sec:schr}). Interestingly, there is no apparent benefit of using SB-SV or SB-VE with a more linear path ($k=0.99$). In fact, PESQ and WhiSQA decrease when using SB-SV with $k=0.99$.  However, compared to SB-SV, the results show that using an exact linear gradient with static variance with ICFM produces higher quality samples. Using ICFM with the FM loss has a marginal improvement over DP. This difference in training objectives suggests that direct gradient estimation is more suitable for ICFM. Together, these findings support the idea that straighter paths are more suitable for flow-based SE, specifically by introducing time-independent variance.
\figureref{fig:N} shows how the number of ODE steps affects the performance of various model types from \tableref{tab:results}. Although SB-VE performs better at 1 ODE step, ICFM requires 20 steps to outperform SB-VE. SB-SV has a similar trend to ICFM, suggesting that static variance reduces performance when using fewer ODE steps. We speculate that models with static variance might perform worse with one-step ODEs because they don't exactly interpolate the data~\eqref{eq:icfm_mean_variance}, unlike SB-VE \eqref{eq:sb_mean_variance}.
On average, the samples of \ac{DDP} are either comparable to or of improved quality over those predicted with 50 ODE steps. Further, possible ODE errors in SB-SV $k=0.99$ are circumvented by DDP. Our proposed ICFM with FM loss reports the highest PESQ and SI-SDR. The results suggest that, although trained for ODE solvers, flow-based models have prominent predictive properties. Another reason ICFM performs well here could be attributed to variance at the boundaries, alleviating potential overfitting caused by exact interpolation.
\vspace{-1.2em}
\section{Conclusion}
 This paper views the time-independence of path gradient and variance as an analogue for straightness. We assessed the impact of probability path straightness on flow-based model performance for SE. By comparing SB-VE with SB-SV, we observed greater improvement with time-independent variance over gradient, but overall found that speech quality metrics were greater improved by using ICFM, which fixes both gradient and variance. However, fixing variance degraded ODE solver performance, but this can be circumvented by directly predicting the data at inference.

\acks{Thanks to Aaron Fletcher for proofreading. Thanks to George Close and Robbie Sutherland for speech enhancement knowledge. This work was supported by the UKRI AI Centre for Doctoral Training in Speech and Language Technologies (SLT) and their Applications funded by UK Research and Innovation [grant number EP/S023062/1]. For the purpose of open access, the author has applied a Creative Commons Attribution (CC BY) licence to any Author Accepted Manuscript version arising.}
\bibliography{pmlr-sample}

\end{document}